\documentclass[prl,twocolumn]{revtex4-1}
\usepackage{amsmath,amssymb,amsfonts,bm,amscd}
\usepackage{mathtools}

\usepackage{graphics}

\makeatletter
\renewcommand{\@makecaption}[2]{
  \vskip\abovecaptionskip
  \sbox\@tempboxa{\small\sf #1: #2}%
  \ifdim \wd\@tempboxa >\hsize
  \small\sf #1: #2\par
  \else
    \global \@minipagefalse
    \hb@xt@\hsize{\hfil\box\@tempboxa\hfil}%
  \fi
  \vskip\belowcaptionskip}
\makeatother

\def\ba{\begin{eqnarray}}
\def\ea{\end{eqnarray}}

\def\frak{\mathfrak}

\def\tilde{\widetilde}

\def\bar{\overline}

\def\Dslash{\,\,{\raise.15ex\hbox{/}\mkern-12mu D}}
\def\Dbarslash{\,\,{\raise.15ex\hbox{/}\mkern-12mu {\bar D}}}
\def\delslash{\,\,{\raise.15ex\hbox{/}\mkern-9mu \partial}}
\def\delbarslash{\,\,{\raise.15ex\hbox{/}\mkern-9mu {\bar\partial}}}
\def\pslash{\,\,{\raise.15ex\hbox{/}\mkern-9mu p}}
\def\calDslash{\,\,{\raise.15ex\hbox{/}\mkern-12mu {\cal D}}}

\newcommand{\C}{{\mathbb C}}

\def\CN{{\mathcal N}}

\def\CR{{\mathcal R}}

\def\CW{{\mathcal W}}

\def\SU{{SU}}
\def\U{{U}}
\def\SO{{SO}}

\newcommand{\cp}{{\mathbb{C}}{\mathbf{P}}}

\renewcommand{\bar}{\overline}

\begin{document}
\preprint{CALT-TH 2017-072}

\title{2d (0,2) appetizer}

\author{Mykola Dedushenko}
\affiliation{\it Walter Burke Institute for Theoretical Physics, California Institute of Technology, Pasadena, CA 91125, USA}

\author{Sergei Gukov}
\affiliation{\it Walter Burke Institute for Theoretical Physics, California Institute of Technology, Pasadena, CA 91125, USA}

\begin{abstract}
Searching for the simplest non-abelian 2d gauge theory with $\CN=(0,2)$ supersymmetry and non-trivial IR physics,
we propose a new duality for $SU(2)$ SQCD with $N_f = 4$ chiral flavors. The chiral algebra of this theory is found to be $\mathfrak{so}(8)_{-2}$, the same as in 4d $\CN=2$ $SU(2)$ gauge theory with four hypermultiplets.
\end{abstract}


\maketitle

\newcommand{\be}{\begin{equation}}
\newcommand{\ee}{\end{equation}}

\section{Introduction}

Two-dimensional theories with $\CN=(0,2)$ supersymmetry play an important role in string theory, quantum field theory,
and connections with pure mathematics.
They describe world-sheet physics of heterotic strings.
In quantum field theory, they can be found on two-dimensional objects (dynamical and non-dynamical)
in a 4d theory with $\CN \ge 1$ supersymmetry.
And many interesting connections with mathematics arise because the BPS sector of a 2d $\CN=(0,2)$ theory
is a chiral CFT (a.k.a. chiral algebra).
Yet, until recently, not much was known about strongly coupled gauge dynamics of 2d $\CN=(0,2)$ theories.

The first exactly solvable example of a 2d $\CN=(0,2)$ SQCD was proposed in \cite{Gadde:2013lxa}
and studied further in \cite{Jia:2014ffa,Gadde:2014ppa,Guo:2015gha,Honda:2015yha,Gadde:2015kda,Gadde:2015wta,Franco:2016nwv,Franco:2016qxh}.
A theory with $N_c$ colors, $N_f$ chiral flavors and $\tilde N_f$ Fermi flavors also contains $2N_c - N_f + \tilde N_f$
anti-fundamental flavors and $N_f (2N_c - N_f + \tilde N_f)$ singlet ``mesons'' that cancel anomalies.
This quite elaborate 2d SQCD exhibits a rich phase diagram, with dynamical SUSY breaking
and a peculiar triality symmetry that mimics symmetries of smooth 4-manifolds (under `handle slides')
and resembles Seiberg duality in four dimensions.
In particular, the triality symmetry permutes $N_f$, $\tilde N_f$, and $2N_c - N_f + \tilde N_f$,
and signals SUSY breaking when these numbers violate triangle inequalities.

In this paper we propose the simplest 2d $(0,2)$ SQCD which, compared to the above class of theories,
is on the edge of dynamical SUSY breaking and, nevertheless, has non-trivial IR physics described by
a $(0,2)$ Landau-Ginzburg (LG) model with a cubic superpotential.
To build this theory we take the simplest non-abelian gauge group $\SU(2)$.
Since 2d $(0,2)$ vector multiplet contains left-moving fermions, we need to cancel gauge anomaly,
which in a theory with $\SU(N_c)$ gauge group and $N_f$ fundamental flavors looks like
\be
\label{noanom}
- N_c + \frac{1}{2} N_f.
\ee
In particular, we see that in a theory with $N_c = 2$ the simplest way to cancel gauge anomaly is to add
$N_f = 4$ chiral multiplets in the fundamental representation of the gauge group.

In the rest of this note we describe the IR dynamics of this theory and propose
a new duality with a $(0,2)$ Landau-Ginzburg model of one Fermi superfield $\Psi$ coupled
to six chiral multiplets $\Phi_i$, $i = 1, \ldots, 6$, via a $(0,2)$ superpotential
\be
J \; = \; \Psi \text{Pf} (\Phi),
\label{LGJ}
\ee
where we find it convenient to think of $\Phi$ as ${\bf 6} \cong \wedge^2 {\bf 4}$
under $\SO(6) \cong \SU(4)$ global symmetry group.

Given a possibility of a dynamical SUSY breaking,
one might feel suspicious or, perhaps, even skeptical about the IR duality between $\SU(2)$ SQCD and the LG model (\ref{LGJ}).
Such concerns are further supported by $c$-extremization \cite{Benini:2012cz},
which leads to negative central charges on both sides of the duality
and usually is a signal for either dynamical SUSY breaking or lack of a normalizable vacuum.
Despite all such indications, we argue that both sides of the proposed duality do not break SUSY and, furthermore, have equivalent IR physics.


\section{Vacua}

Let us start by analyzing the space of classical vacua.
On the gauge theory side, the classical space of vacua is a ``complex cone'' on the Grassmannian
\be
\text{Gr} (2,4) = V^{\otimes 4} /\!/ \U(2),
\label{Grquotient}
\ee
where $V \cong \C^2$ denotes the fundamental representation of $\U(2)$.
One way to see this is to imagine (formally) that our gauge group is $\U(2)$ rather than $\SU(2)$.
Then, the space of vacua in such fictitious theory would be the K\"ahler quotient (\ref{Grquotient}).
Going back to our original theory, that is replacing the gauge group $\U(2)$ by $\SU(2)$,
in (\ref{Grquotient}) has the effect of removing the D-term constraint and a quotient by $\U(1)$.
Hence, the resulting moduli space of vacua is a complex cone on $\text{Gr} (2,4)$.

The same moduli space of classical vacua can be easily identified on the Landau-Ginzburg side.
Indeed, the same Grassmannian (\ref{Grquotient}) in Pl\"ucker coordinates is a hypersurface in $\cp^5$:
\be
\text{Gr} (2,4) = \{ \Phi_{12} \Phi_{34} - \Phi_{13} \Phi_{24} + \Phi_{23} \Phi_{14} = 0 \}.
\ee
Since $\cp^5 = \mathbb{C}^6 /\!/ \U(1)$ is a K\"ahler quotient, just as on the gauge theory side
discussed earlier, we can ``ungauge the $\U(1)$'' by removing the $\U(1)$ quotient and the moment map
constraint associated with it. The result, of course, is simply the hypersurface in $\C^6$:
\be
\Phi_{12} \Phi_{34} - \Phi_{13} \Phi_{24} + \Phi_{23} \Phi_{14} = 0.
\label{PluckerM}
\ee
This is the same five-complex-dimensional space of vacua we found in $\SU(2)$ SQCD with $N_f = 4$.
According to a theorem of Tian and Yau \cite{TianYau}, this quadric admits a cohomogeneity one Ricci-flat metric
that comes in a family of non-compact Calabi-Yau 5-folds obtained by deforming the complex structure in (\ref{PluckerM}):
\be
\Phi_{12} \Phi_{34} - \Phi_{13} \Phi_{24} + \Phi_{23} \Phi_{14} = \epsilon.
\label{PfCY}
\ee
The conical singularity at $\epsilon =0$ is at a finite distance in moduli space \cite{Gukov:1999ya}.

In quantum theory, there are several possible scenarios for the physics associated with this singularity.
For instance, one option is that the conical singularity at $\Phi=0$ is simply resolved in quantum theory, {\it cf.} \cite{Aharony:2016jki}.
Another option is that the singularity is hiding extra degrees of freedom and / or another branch of vacua emanating from it.

Note, the Pfaffian Calabi-Yau (\ref{PfCY}) is a non-compact analogue of a 2d $\CN=(2,2)$ model considered in \cite{Hori:2006dk}.


\section{Compactification from 4d}

An insight into quantum physics of the proposed dual pair of 2d $(0,2)$ theories comes from connecting them to
another dual pair, namely 4d $\CN=1$ gauge theories related by Seiberg duality.

Seiberg duality \cite{Seiberg:1994pq} relates 4d $\CN=1$ SQCD with gauge group $\SU(N_c)$ and $N_f$ fundamental flavors $q_i$
to a similar theory with $\SU(N_f - N_c)$ gauge group, $N_f$ fundamental flavors, singlet ``mesons'' $M_i{}^j \sim q_i \, \tilde{q}^j$,
and a certain superpotential.
As we explain in the rest of this section, a special instance of this 4d duality,
when compactified on a 2-sphere with a partial topological twist, gives precisely the proposed dual pair
of 2d $\CN=(0,2)$ SQCD and the Landau-Ginzburg model with the Pfaffian superpotential (\ref{LGJ}).

To describe topological reduction {\it a la} \cite{Johansen:1994ud, Bershadsky:1995vm, Closset:2013sxa, Gadde:2015wta} of a 4d $\CN=1$ theory on $S^2$ --- or,
more generally, on a genus $g$ Riemann surface $F_g$ --- we need to describe what happens to three basic ingredients:
$\CN=1$ vector multiplets, chiral multiplets, and the superpotential interaction.
We start with a vector multiplet, whose topological reduction on a genus $g$ Riemann surface is relatively simple:
it gives one 2d $(0,2)$ vector multiplet (with the same gauge group) and $g$ adjoint $(0,2)$ chiral multiplets,
see {\it e.g.} \cite{Gukov:2017zao}:
$$
\text{4d } \CN=1 \text{ vector}
\quad \xrightarrow[~]{~\text{on $F_g$}~} \quad
\begin{cases}
\text{2d } \CN = (0,2) \text{ vector}, & \\
+~g~ \CN = (0,2) \text{ chirals}. &
\end{cases}
$$

Topological reduction of 4d $\CN=1$ chiral multiplets is more subtle and depends on the choice of the $R$-symmetry,
under which all chiral fields must carry integer $R$-charges.
Of course, the $R$-symmetry should also be non-anomalous, {\it e.g.} for $\SU(N_c)$ theory with $N_f$ fundamental flavors
the $R$-charges of the chiral multiplets must satisfy
\be
\sum_{i = 1}^{2N_f} R_i \; = \; 2 (N_f - N_c).
\label{4danomaly}
\ee
Assuming these conditions are met, the topological reduction of a 4d $\CN=1$ chiral multiplet with
$R$-charge $R$ yields 2d spectrum of $(0,2)$ chiral and Fermi multiplets controlled by
the following cohomology groups:
\be
\text{2d } \CN=(0,2)~
\begin{cases}
H^0 (K^{R/2} \otimes L({\frak m})) \text{ chirals}, & \\
H^0 (K^{1 - R/2} \otimes L(-{\frak m})) \text{ Fermi}, &
\end{cases}
\ee
where $K$ is the canonical bundle of $F_g$ and we also allowed a coupling to general background flux ${\frak m}$.
In particular, for a genus-0 compactification on $S^2$, each 4d chiral multiplet contributes
to the 2d $\CN=(0,2)$ field content either $1-R$ chirals if $R<1$ or $R-1$ Fermi if $R>1$, {\it cf.} \cite{Gadde:2015wta}.

Finally, the superpotential terms in four dimensions, upon the topological reduction on $F_g$,
yield superpotential $E$ or $J$ terms of the effective $\CN=(0,2)$ theory in two dimensions.

Now, let us apply these simple rules to 4d $\CN=1$ SQCD with $N_c = 2$ and $N_f = 3$.
Note, these numbers of colors and flavors obey $N_f = N_c + 1$ and also $N_f = \frac{3}{2} N_c$.
Since $N_f = 3$, this theory has a total of 6 chiral multiplets that transform as doublets under the $\SU(2)$ gauge group.
We choose the following assignment of integer $R$-charges, which satisfy the anomaly cancellation condition (\ref{4danomaly}):
\be
R \; = \; (1,1,0,0,0,0).
\label{Rchoice}
\ee
With these $R$-charge assignments, the spectrum of the theory on $S^2$ consists of one $\SU(2)$ vector multiplet and four fundamental chirals.
In other words, it is precisely our candidate for the simplest non-abelian 2d $\CN=(0,2)$ gauge theory with non-trivial IR physics.

In a similar way, we can also derive the Landau-Ginzburg model (\ref{LGJ}) from the topological reduction of
the Seiberg dual to 4d $\CN=1$ SQCD with $N_c =2$ and $N_f = 3$.
Since $N_f - N_c = 1$, the dual theory is a Landau-Ginzburg model already in four dimensions.
It has 15 ``meson'' fields $M_{ij}$ interacting via a cubic superpotential:
\be
\CW \; = \; \text{Pf} (M).
\label{Wsuper}
\ee
The $R$-charges of $M_{ij}$ compatible with (\ref{Rchoice}) can be easily deduced from
the relation $M_{ij} \sim \epsilon_{ab} \, q_i^a \, q_j^b$,
where $q_i^a$ are the fundamental ``quarks'' of the original 4d $\CN=1$ SQCD.
Specifically, we have
\be
R (M_{12}) \; = \; 2,
\ee
\be
R(M_{1 i \ne 2}) = R (M_{2 j \ne 1}) = 1,
\ee
and the six components $M_{ij}$ with $i$ and $j$ not equal to 1 or 2 all have $R (M_{ij}) = 0$.
Upon topological reduction on a 2-sphere, the latter give rise to 2d $\CN=(0,2)$ chiral multiplets $\Phi_i$, $i = 1, \ldots, 6$.
The mesons with $R=1$ do not contribute to the spectrum of 2d theory at all,
whereas the component $M_{12}$ gives rise to a Fermi multiplets $\Psi$.
Moreover, the 4d superpotential interaction (\ref{Wsuper}) reduces to the 2d Pfaffian superpotential (\ref{LGJ}).
All in all, this is precisely the Landau-Ginzburg model that was proposed as IR dual to 2d $\CN=(0,2)$ SQCD.
Here, we related the proposed IR duality in two dimensions to a more familiar Seiberg duality in 4d.

As further evidence for the proposed duality, one can compare elliptic genera
of 2d $\CN=(0,2)$ SQCD and the Landau-Ginzburg model using the theta function identity (4.7) from \cite{Putrov:2015jpa}. In the next section we match even stronger invariants on both sides of the duality.

\section{The chiral algebra}

2d $(0,2)$ theories are known to have a sector of BPS operators, defined by passing to the cohomology of $\bar{Q}_+$, whose OPE has a structure of a vertex operator algebra, or chiral algebra \cite{Witten:1993jg, Silverstein:1994ih}. The exact chiral algebra is known to be an RG invariant \cite{Dedushenko:2015opz}, hence it can serve as a useful check of dualities, whenever it is possible to compute it. This philosophy was utilized in the literature before \cite{Guo:2015gha} mainly through the study of chiral rings (regular subsectors in chiral algebras) in $(0,2)$ NLSMs \cite{Katz:2004nn, Guffin:2008pi, Melnikov:2009nh}, which can be applied to gauge theories through their low-energy sigma model descriptions.

Here we are going to apply the chiral algebra machinery to our gauge theory and its Landau-Ginzburg dual directly, rather than through the looking glass of NLSMs. We begin on the gauge theory side, explaining the method from \cite{Notes} along the way. First of all, we work under the hypothesis that the algebra is perturbatively exact. While instantons are known to sometimes drastically change the answer \cite{Witten:2005px, Nekrasov:2005wg, Tan:2008mi, Yagi:2010tp}, we believe it does not happen in our dual pair. The theory at hands admits no familiar vortex solutions, and moreover, results on the Landau-Ginzburg side confirm our assumption. 

On the gauge theory side, the chiral algebra can be computed in perturbation theory in the gauge coupling $e$, which is essentially the spectral sequences method applied to the $\bar{Q}_+$ cohomology. In the zeroth order in $e$, we simply impose the Gauss law constraint on the product of chiral algebras of free multiplets. The free $(0,2)$ chiral multiplet (valued in a representation $\CR$) contributes a $\beta\gamma$ system of conformal weights $(h_\beta, h_\gamma)=(1-\lambda, \lambda)$ (valued in the same representation), which we denote $(\beta,\gamma)^{(\lambda)}$. Parameter $\lambda$ is not fixed and is related to the true R-symmetry of the IR CFT. The chiral algebra of a free vector multiplet is given by a small algebra of the $bc$ ghost system valued in the adjoint of the gauge group; we call it $(b,c)_{\rm small}$. Its conformal dimensions are $(h_b, h_c)= (1,0)$, which are not ambiguous because vector multiplet has canonical R-charge. The term ``small algebra'' means that operators constructed from basic building blocks $b(z)$ and $c(z)$ can only contain derivatives of $c(z)$ but not $c(z)$ itself. So the zeroth order approximation to the chiral algebra is given, after imposing gauge invariance, by:
\begin{equation}
E_0 = \left((\beta, \gamma)^{(\lambda)} \otimes (b, c)_{\rm small}\right)^G.
\end{equation}
We then include all higher-order correction to $\bar{Q}_+$ and see that the way it acts on $E_0$ coincides precisely with the BRST operator. Namely, the BRST current is:
\begin{align}
J_{BRST} &= \sum_{A=1}^{\dim G} c^A (J^A_{\rm m} + \frac12 J^A_{\rm gh}),\cr 
J^A_{\rm m} &= i\beta T^A\gamma,\quad J^A_{\rm gh}=-i f^A{}_{BC}c^A b^B c^C,
\end{align}
and it defines a $Q_{BRST}$ operator in the usual way. One can check that it is nilpotent, $Q_{BRST}^2=0$, precisely when the anomaly cancellation condition \eqref{noanom} holds, so it is consistent to study its cohomology. The perturbative chiral algebra is then simply given by a BRST reduction of the ``ungauged'' chiral algebra \cite{Notes}:
\begin{equation}
H_{\rm pert}(\bar{Q}_+) \cong H_{\rm BRST}(E_0).
\end{equation}

Those familiar with the work of \cite{Beem:2013sza} may find a striking similarity between this procedure and the one for the chiral algebra of Lagrangian $4d$ $\CN=2$ SCFTs. Their prescription is given by a BRST reduction of the symplectic boson valued in the matter representation. Note that symplectic boson CFT is the same thing as the $\beta\gamma$ system at $\lambda=\frac12$ (this $\lambda$ is different from the one in \cite{Fluder:2017oxm}, where it referred to twisted sectors). Moreover, the BRST cohomology problem does not depend on the value of $\lambda$. Therefore, our prescription is not just similar to \cite{Beem:2013sza}, it is exactly the same. 

This simple observation, whether it has any deep meaning or not, helps us to avoid a lot of technicalities. It shows that our chiral algebra for 2d $\SU(2)$ gauge theory with four chirals coincides with the chiral algebra of $4d$ $\CN=2$ $\SU(2)$ gauge theory with $N_f=4$. There was a lot of evidence in \cite{Beem:2013sza} that the latter is given by an $\mathfrak{so}(8)$ current algebra at level $-2$, so our answer is:
\begin{equation}
H_{\rm pert}(\bar{Q}_+) \cong \mathfrak{so}(8)_{-2}.
\end{equation}
Even though this algebra does not depend on parameter $\lambda$, its value does not get washed out completely. It enters in the choice of the stress-energy tensor, which coincides with the physical left-moving stress-energy tensor of the IR CFT. Only for $\lambda=\frac12$ it would coincide with the Sugawara tensor. For different values of $\lambda$, we have:
\begin{equation}
T = T_{\rm Sug} + \Delta T,\quad \Delta T = (1/2 -\lambda)\partial(\beta\gamma).
\end{equation}

Now we would like to approach this problem from the Landau-Ginzburg side, where we use the methods of \cite{Dedushenko:2015opz}. To compute the cohomology of $\bar{Q}_+$, we identify the cohomology of a superspace derivative $\bar{D}_+$ acting on superfields and then take their lowest components. One has to take into account the operator equations of motion:
\begin{align}
\bar{D}_+ \partial_{--} \bar\Phi^{ij} &= -i\Psi \varepsilon^{ijkl}\Phi_{kl},\cr
\bar{D}_+ \bar\Psi &= -2 {\rm Pf}(\Phi),
\end{align}
where $\partial_{--}$ equals the holomorphic derivative upon Wick rotation. Since $\Phi_{ij}$ and $\Psi$ are annihilated by $\bar{D}_+$, they are in the cohomology, and equations of motion imply relations ${\rm Pf}(\Phi)=0$ and $\Psi \Phi_{ij}=0$ in the cohomology. Assuming that the fields $\Phi_{ij}$ have R-charge $\alpha$ in the IR, we can further identify the stress-energy tensor in the cohomology:
\begin{align}
T=\sum_{i>j}\left[ \partial_{--}\Phi_{ij}\partial_{--}\bar\Phi^{ij} - \frac{\alpha}2 \partial_{--}(\Phi_{ij}\partial_{--}\bar\Phi^{ij})\right]\cr
+\frac{i}2 \Psi\partial_{--}\bar\Psi -\frac{i}{2} \partial_{--}\Psi \bar\Psi + \frac{i(1-2\alpha)}{2} \partial_{--}(\Psi\bar\Psi),
\end{align}
as well as $\mathfrak{u}(4)$ currents:
\begin{equation}
J_i{}^j = i\Phi_{ik}\partial_{--}\bar\Phi^{kj} + \delta_i^j \bar\Psi \Psi.
\end{equation}
Composite operators should be defined with care: we always have to subtract singularities appearing in colliding their constituent elementary fields. This cannot lift $T$ and $J_i{}^j$ from the cohomology because their singularities only have numeric coefficients, but can affect more complicated composite operators. We can identify such composite operator in the classical $\bar{D}_+$ cohomology:
\begin{align}
2i\bar\Psi \Psi \partial_{--}\bar\Phi^{ij} - \partial_{--}\bar\Phi^{ni}\Phi_{np}\partial_{--}\bar\Phi^{pj}.
\end{align}
To properly define quantum operators, we have to renormalize them by subtracting singularities. The second term is actually non-singular, but we still need to split points in order to correctly evaluate the action of $\bar{D}_+$ on it. We define renormalized operators:
\begin{align}
&[\bar\Psi \Psi \partial_{--}\bar\Phi^{ij}]_{\rm ren}(0)\cr 
&\quad\quad\quad= \lim_{\epsilon\to 0}(\bar\Psi(\epsilon) \Psi(0) \partial_{--}\bar\Phi^{ij}(0) - \frac{i}{\epsilon} \partial_{--}\bar\Phi^{ij}(0)),\cr
&[\partial_{--}\bar\Phi^{ni}\Phi_{np}\partial_{--}\bar\Phi^{pj}]_{\rm ren}(0) \cr 
&\quad\quad\quad=\lim_{\epsilon\to 0} \partial_{--}\bar\Phi^{ni}(0)\Phi_{np}(\epsilon)\partial_{--}\bar\Phi^{pj}(2\epsilon).
\end{align}
A very accurate computation at finite $\epsilon$ shows that now:
\begin{align}
\bar{D}_+\left\{ 2i[\bar\Psi \Psi \partial_{--}\bar\Phi^{ij}]_{\rm ren} - [\partial_{--}\bar\Phi^{ni}\Phi_{np}\partial_{--}\bar\Phi^{pj}]_{\rm ren} \right\}\cr
=4\partial_{--}(-i\Psi\varepsilon^{ijkl}\Phi_{kl}).
\end{align}
In an analogous situation in \cite{Dedushenko:2015opz}, similar computation was used to argue that the cohomology class was lifted at the quantum level. In our case, though, the right-hand side is actually $\bar{D}_+$-exact due to the equation of motion, being equal to $4 \bar{D}_+ \partial_{--}^2 \bar\Phi^{ij}$. We conclude that in our case, quantum effects renormalize the cohomology class, and the correct one is given by:
\begin{align}
B^{ij}&=2i[\bar\Psi \Psi \partial_{--}\bar\Phi^{ij}]_{\rm ren}\cr 
&- [\partial_{--}\bar\Phi^{ni}\Phi_{np}\partial_{--}\bar\Phi^{pj}]_{\rm ren} - 4\partial_{--}^2\bar\Phi^{ij}.
\end{align}

This composite operator groups together with $\Phi_{ij}$ and $J_i{}^j$ to form the $\mathfrak{so}(8)_{-2}$ current algebra, precisely matching the dual gauge theory result, including the OPE. One could call this an accident in the terminology of \cite{Bertolini:2014ela}. Denoting $\gamma_i\cdot \gamma_j = \varepsilon^{\alpha\beta} \gamma_{i\alpha}\gamma_{j\beta}$, where $\alpha,\beta$ are indices in the fundamental of $\mathfrak{su}(2)$, we have:
\begin{align}
\gamma_i\cdot \gamma_j &\leftrightarrow \Phi_{ij},\cr
\beta^i\cdot \gamma_j &\leftrightarrow \partial_{--}\bar\Phi^{ik}\Phi_{kj} - i\delta^i_j \bar\Psi \Psi,\cr
\beta^i\cdot \beta^j &\leftrightarrow B^{ij},\cr
4\lambda &= \alpha.
\end{align}
The Landau-Ginzburg side has an extra fermionic operator $\Psi$ in the cohomology, which might look puzzling. However, our duality suggests that it simply decouples as a free field along the flow. The way it enters the algebra shows that it is completely consistent to make such a truncation, so that the interesting part of the chiral algebra is indeed $\mathfrak{so}(8)_{-2}$. We can also conjecture that there are no other independent composite operators that have to be taken into account.

\section{Concluding remarks}
Some of the facts observed in this note call for further investigation. In particular, the match of our chiral algebra with the one from \cite{Beem:2013sza} sounds intriguing and raises a question whether it is a pure coincidence or a manifestation of some deeper connection.

But most importantly, a lot of standard techniques are not applicable to our theory due to the lack of normalizable vacuum. One example is the $c_R$-extremization \cite{Benini:2012cz}, equivalent to $c_L$-extremization (due to $c_L-c_R$ being fixed by the gravitational anomaly, equal to $-5$ in our case), which in turn is equivalent to the requirement that all $U(1)$ currents are primary \cite{Dedushenko:2015opz}. Applying it to our theory gives $\lambda=1/2$, $c_L=-14$. This is not in conflict with unitarity because of the lack of normalizble vacuum. But for the same reason, this value of $c_L$ does not have to be correct because $c$-extremization also fails. On top of that we should add that in a theory with normalizable vacuum, decoupling of $\Psi$ along the flow would immediately imply that its dimension is at the unitary bound. This does not have to hold in our case either. Overall, we are lacking one extra handle on the dynamics of our model to say more about its IR physics.

Somewhat related, the fate of singularity at the origin of the moduli space is not entirely clear as well. For this puzzle, we can make a guess: if topological reduction on $S^2$ from 4d to 2d commutes with the RG flow, we can simply look at the $\CN=1$ $\SU(2)$ gauge theory in 4d with $N_f=3$, which is a parent of our 2d theory. According to \cite{Seiberg:1994bz}, the classical moduli space in such a theory does not receive quantum corrections, and singularity at the origin carries some massless degrees of freedom, so we could guess that the same happens in our case. Furthermore, the relation ${\rm Pf}(\Phi)=0$ in the chiral algebra gives further evidence for this claim. Finally, we should note that this problem might be amenable to the methods of \cite{Beasley:2004ys} extended to 2d gauge theories. 






\begin{acknowledgements}
We thank M.Fluder, J.Heckman, D.Jafferis, D.Kutasov, N.Nekrasov, P.Putrov, J.Song for useful discussions. This work was supported by the Walter Burke Institute for Theoretical Physics and the U.S. Department of Energy, Office of Science, Office of High Energy Physics, under Award No.\ DE{-}SC0011632. The work of MD was also supported by the Sherman Fairchild Foundation. SG gratefully acknowledges support from Harvard University, where some of the research for this  paper  was  performed  during  the fall 2017, as well as partial support by the National Science Foundation under Grant No. NSF PHY11-25915 and Grant No. NSF DMS 1664240.
\end{acknowledgements}

\end{document}